\documentclass[a4paper,notitlepage,12pt]{article}

\begin{document}
\title{Steady thermocapillary migration of a droplet in a uniform temperature gradient combined with a radiation energy source at large Marangoni numbers}
\author{ Zuo-Bing Wu$^{1,2}$\footnotemark[1]\\
 $^1$State Key Laboratory of Nonlinear Mechanics,
 Institute of Mechanics,\\
  Chinese Academy of Sciences, Beijing 100190, China\\
  $^2$School of Engineering Science, \\University of Chinese Academy of Sciences,
  Beijing 100049, China}
 \maketitle

\footnotetext[1]{Corresponding author. Tel:. +86-10-82543955;
fax.: +86-10-82543977. \\
Email addresses: wuzb@lnm.imech.ac.cn (Z.-B. Wu)}

\newpage
\begin{abstract}
The steady thermocapillary droplet migration in a uniform temperature gradient combined with
a radiation energy source at large Reynolds and Marangoni numbers is studied.
To reach a terminal quasi-steady process, the magnitude of the radiation energy source
is required to preserve the conservative integral thermal flux across the surface.
Under the quasi-steady state assumption, an analytical result for the steady thermocapillary migration
of a droplet at large Reynolds and Marangoni numbers is derived by using the method
of matched asymptotic expansions. It is shown
that the thermocapillary droplet migration speed
increases as Marangoni number increases, while the radiation energy source
with the sine square dependence is provided.


\textbf{Keywords} \ Interfacial tension; Thermocapillary
migration of droplet; Quasi-steady state assumption; Large Marangoni number; Microgravity\\
\end{abstract}

\newpage
\section{Introduction}
A droplet in an external fluid or on a solid substrate can be driven by
body forces generated in the gravitational, electric, magnetic, and ultrasonic fields \cite{0}.
Even in the absence of the body forces, the variable surface tension along the interface
can also drive the droplet migration in the external fluid or the solid substrate.
Thermocapillary migration of a droplet in microgravity environment is a very interesting topic on both fundamental
hydrodynamic theory and engineering application \cite{1}.
Young et al \cite{2} carried out an initial
study on thermocapillary migration of a droplet in a uniform temperature
gradient in the limits of zero Reynolds(Re) and Marangoni(Ma) numbers (YGB model).
Subramanian \cite{4} proposed the quasi-steady state assumption
and obtained analytical results with high order expansions at small Ma numbers.
The thermocapillary droplet migration processes at small Ma numbers are understood
very well in the series of theoretical analyses, numerical simulations and
experimental investigations \cite{5,6}. However,
the physical behaviors at large Ma numbers appear rather complicated due to
the momentum and energy transfer though the interface of two-phase fluids.
Meanwhile, to perform a feasible numerical simulation of thermocapillary migration of a droplet
at large Ma number is still a challenge due to very thin thermal boundary [$O(Ma^{-1/2})$]
and very long migration time [$O(Ma)$].
Under the assumption of the quasi-steady state, Balasubramaniam \& Subramanian reported \cite{7}
that the migration speed of a droplet
increases with the increasing of Ma number, as is in qualitative
agreement with the corresponding numerical simulation \cite{8}.
The experimental investigation carried out by
Hadland et al \cite{9} and Xie et al \cite{10} shown that the droplet migration speed
 decreases as Ma number increases, which is
qualitatively discrepant from the above theoretical and numerical
results.
Wu \& Hu \cite{11} and Wu \cite{12} identified a nonconservative integral thermal
flux across the surface in the steady thermocapillary droplet migration
at large Ma(Re) numbers, which indicates that the thermocapillary
droplet migration at large Ma(Re) numbers is an unsteady process.
To preserve a conservative integral thermal flux across the surface,
two methods, i.e., the adding thermal source inside the droplet or at the surface,
were also suggested.
With a thermal source added inside the droplet,
an analytical result of the steady thermocapillary migration of the droplet
 at large Ma(Re) numbers was determined \cite{13}.
Therefore, the thermocapillary droplet migration at large Ma numbers remains
a topic to be studied with respect to its physical mechanism.

In the above studies, the variable surface
tension exerted on the interface of two-phases is generated by adding
a non-uniform temperature field. On the other hand, a radiative heating in contrary to the direction of movement,
which provides a thermal source at the surface through the absorption, can also form
 the variable surface tension exerted on the interface.
Oliver \& Dewitt \cite{3} firstly analyzed thermocapillary migration of a droplet
caused by a thermal radiation in microgravity environment in the zero Re and Ma number limits.
Rednikov \& Ryzzantsev \cite{13a} independently derived the similar results and determined
the deformation of the droplet.
Zhang \& Khodadadi \cite{14a} and Khodadadi \& Zhang \cite{14} numerically studied
the effects of thermocapillary convection on melting of droplets at
a short-duration and an uniform heat pulses under zero gravity conditions at large Ma numbers, respectively.
Lopez et al \cite{15} experimentally observed the thermocapillary migration of a droplet caused
by a laser beam heating due to the absorption of the laser radiation
in making a strongly non-homogeneous distribution of
temperature inside the droplet as well as at its surface.

 In this paper, a radiation energy source in contrary to the direction of movement is placed
 to preserve the conservative integral thermal
 flux across the surface, thermocapillary droplet migration
 at large Re and Ma numbers can thus reach a quasi-steady process.
The steady thermocapillary droplet migration in the uniform temperature gradient combined with
the radiation energy source at large Re and Ma numbers is studied.
In comparing with the previous method to preserve a conservative integral thermal flux across the surface \cite{13},
the current method, i.e., placing the radiation energy source at the outside of the droplet,
is easier to carry out a real space experiment. In the principle, the previous method adds a thermal
source in the energy equation within the droplet, but the current method adds a heat flux at the interface
of the droplet. The paper is organized as follows.
In Sect. 2,
the magnitude of the radiation energy source is required to preserve the conservative integral thermal
flux across the surface. An analytical result for the steady thermocapillary droplet migration
at large Re and Ma numbers is determined in Sect. 3. Finally, in Sect. 4, the conclusions and discussions are given.

\section{Problem formulation}
Consider
the thermocapillary migration of a spherical droplet of radius $R_0$,
density $\gamma \rho$, dynamic viscosity $\alpha \mu$, thermal
conductivity $\beta k$, and thermal diffusivity $\lambda \kappa$
in a continuous phase fluid of infinite extent with density
$\rho$, dynamic viscosity $\mu$, thermal conductivity $k$, and
thermal diffusivity $\kappa$ under a uniform temperature gradient
$G$ in the direction of movement and an inhomogeneous radiation energy source $S$
in contrary to the direction of movement.
 It is assumed that
the continuous phase fluid is transparent and that
the radiation is absorbed totally on the droplet surface.
The rate of change of the interfacial tension between the
droplet and the continuous phase fluid with temperature is denoted by
$\sigma_T$. Unsteady energy equations for the continuous phase
and the fluid in the droplet in a laboratory
coordinate system denoted by a bar are written as follows
\begin{equation}
\begin{array}{l}
\frac{\partial{\bar{T}}}{\partial t} + \bar{\bf v} \bar{\nabla} \bar{T}= \kappa \bar{\Delta} \bar{T},\\
\frac{\partial{\bar{T'}}}{\partial t} + \bar{\bf v'} \bar{\nabla} \bar{T'}= \lambda \kappa \bar{\Delta} \bar{T'},
\end{array}
\end{equation}
where $\bar{\bf v}$ and $\bar{T}$ are velocity and temperature,
a prime denotes quantities in the droplet.
The solutions of
Eqs. (1) have to satisfy the boundary conditions at infinity
\begin{equation}
\bar {T} \rightarrow T_0 + G\bar {z},
\end{equation}
where $T_0$ is the undisturbed temperature of the continuous phase
and the boundary conditions at the interface ${\bar {\bf r}}_b$ of the two-phase fluids
\begin{equation}
\begin{array}{l}
\bar T({\bar {\bf r}}_b,t) =\bar T'({\bar {\bf r}}_b,t),\\
\frac{\partial{\bar T}}{\partial n}({\bar {\bf r}}_b,t) + S = \beta
\frac{\partial{\bar T'}}{\partial n}({\bar {\bf r}}_b,t).
\end{array}
\end{equation}

Under the quasi-steady state assumptions, steady axisymmetric energy equations
 non-dimensionalized by taking the radius of the droplet
$R_0$, the YGB model velocity $v_o=-\sigma_T G R_0/\mu$ and $GR_0$ as reference quantities to
make coordinates, velocity and temperature dimensionless
can be written in the spherical coordinate system ($r, \theta$)
 moving with the droplet velocity $V_{\infty}$
as follows
\begin{equation}
\begin{array}{l}
1+u \frac{\partial{T}}{\partial{r}} +\frac{v}{r} \frac{\partial{T}}{\partial{\theta}}= \epsilon^2 \Delta T,\\
1+u' \frac{\partial{T'}}{\partial{r}} +\frac{v'}{r} \frac{\partial{T'}}{\partial{\theta}}=\lambda \epsilon^2 \Delta T',
\label{1}
\end{array}
\end{equation}
where the small parameter $\epsilon$ and Ma number are
defined, respectively, as
\begin{equation}
\epsilon=\frac{1}{\sqrt{Ma V_{\infty}}}
\end{equation}
and
\begin{equation}
Ma=\frac{v_0R_0}{\kappa}.
\end{equation}
The boundary conditions (2)(3) are rewritten, respectively, as
\begin{equation}
T \to r \cos \theta, {\rm as} \  r \to \infty
\end{equation}
and at the interface of two phase fluids
\begin{equation}
\begin{array}{l}
T(1,\theta) =T'(1,\theta),\\
\frac{\partial{T}}{\partial r}(1,\theta) + \Omega \sin^2 \theta \cos \theta = \beta \frac{\partial{T'}}{\partial r}(1,\theta), \ \ \ 0 \leq \theta \leq \pi/2,\\
\frac{\partial{T}}{\partial r}(1,\theta) = \beta \frac{\partial{T'}}{\partial r}(1,\theta), \ \ \ \pi/2 < \theta \leq \pi.
\end{array}
\end{equation}
The inhomogeneous radiation energy source non-dimensionalized by the reference quantity $kG$ is assumed as $S=\Omega \sin^2 \theta$.
Its contribution to the interface thermal flux $S\cos \theta$ is zero at $\theta=\pi/2$, which reveals that the upper and lower interface thermal boundary conditions in Eq. (8) are continuous.
A schematic diagram of the thermocapillary droplet migration in
a coordinate system moving with the droplet speed $V_\infty$ is shown in Fig. 1.

For large Re numbers ($Re=\frac{v_0 R_0}{\nu}$), the velocity fields of the continuous phase and
the fluid within the droplet can be described by potential flows and boundary layers flows, as shown in Fig. 2.
The scaled potential flow fields around a fluid sphere
\begin{equation}
\begin{array}{l}
u=- \cos \theta (1-\frac{1}{r^3}),\\
v= \sin \theta (1+\frac{1}{2r^3})
\end{array}
\end{equation}
and

\begin{equation}
\begin{array}{l}
u'= \frac{3}{2} \cos \theta (1-r^2),\\
v'=- \frac{3}{2} \sin \theta (1-2r^2)
\end{array}
\end{equation}
are taken as those in the continuous
phase and within the droplet, respectively \cite{17,19}.
It is noticed that the potential flow fields (9)(10) for large Re numbers may be obtained from the general solutions
for small Re numbers by setting $D_n=0, n \geq 3$ \cite{20,21}.
For large Ma numbers, the temperature field at infinity in Eq. (7) is further expressed as \cite{11}
\begin{equation}
T \approx r \cos \theta -\frac{1}{2r^2} \cos \theta +o(1).
\end{equation}

Integrating Eqs. (4) in the continuous phase domain
$(r\in [1,r_{\infty}],\theta\in[0,\pi])$
 with the boundary condition (11) and within
the droplet region $(r\in [0,1],\theta\in[0,\pi])$, respectively, we obtain

\begin{equation}
\int_0^{\pi} \frac{\partial{T}}{\partial{r}} (1,\theta) \sin \theta d \theta
+ \int_0^{\pi/2} \Omega \sin^3 \theta \cos \theta d \theta
= -\frac{1}{3 \epsilon^2} + \frac{\Omega}{4}\\
\end{equation}
and

\begin{equation}
\int_0^{\pi} \frac{\partial{T'}}{\partial{r}} (1,\theta) \sin
\theta d \theta = \frac{2}{3 \lambda \epsilon^2}.
\end{equation}
From Eq. (12) and Eq. (13), we have
\begin{equation}
\begin{array}{l}
\beta \int_0^{\pi} \frac{\partial{T'}}{\partial{r}}
(1,\theta) \sin \theta d \theta - \int_0^{\pi}
\frac{\partial{T}}{\partial{r}} (1,\theta) \sin \theta d
\theta - \int_0^{\pi/2} \Omega \sin^3 \theta \cos \theta d \theta\\
= \frac{1}{3 \epsilon^2}
(1+ \frac{2 \beta}{\lambda}) - \frac{\Omega}{4}.
\end{array}
\end{equation}
 For large Ma numbers and finite $V_\infty$,
 Eqs. (12) and (13) should satisfy the thermal flux boundary condition (8),
 i.e., the right side of Eq. (14) will be zero. So, we have
\begin{equation}
\Omega = \frac{4}{3 \epsilon^2} (1 + \frac{2 \beta}{\lambda})
= \frac{4}{3} (1 + \frac{2 \beta}{\lambda}) V_\infty Ma,
\end{equation}
which preserves the conservative integral thermal flux across the surface.
In following, we will focus on the
steady thermocapillary migration of a droplet in the uniform temperature gradient $G$ combined with the external thermal radiation source $S$
and determine the dependence of the migration speed on large Ma number.

\section{Analysis and results}
\subsection{Outer temperature field in the continuous phase} By
using an outer expansion for the scaled temperature field in the
continuous phase
\begin{equation}
T =T_0 +\epsilon T_1 + o(\epsilon),
\end{equation}
the energy equation for the outer temperature field in its leading
order can be obtained from Eqs.(4) as follows
\begin{eqnarray}
1+u \frac{\partial{T_0}}{\partial{r}} +\frac{v}{r} \frac{\partial{T_0}}{\partial{\theta}}= 0.
\end{eqnarray}
By using the coordinate transformation from $(r,
\theta)$ to $(\psi,\theta)$ in the solving Eq. (17), its solution can be written as
\begin{equation}
T_0(r,\theta) = G(\psi) - \int \frac{2r^4}{2r^3+1} \frac{d \theta}{\sin \theta},
\end{equation}
where $G(\psi)$ is a function of $\psi$ (the stream function in the
continuous phase).
Following \cite{7,13}, the solution near $r=1$ is simplified as
\begin{equation}
\begin{array}{ll}
T_0(r,\theta) =& (1+\frac{\pi}{6\sqrt{3}}-\frac{1}{6} \ln432) -\frac{1}{18} (\frac{\pi}{\sqrt{3}} +\ln432)
(r^2-\frac{1}{r}) \sin^2 \theta + \frac{1}{3} \ln(r^2-\frac{1}{r})\\
 & +\frac{2}{3}\ln(1+\cos \theta)
+ \frac{2}{9} (r^2 -\frac{1}{r}) \cos \theta +\frac{1}{9}  (r^2 -\frac{1}{r}) \ln(r^2 -\frac{1}{r}) \sin^2 \theta\\
& +\frac{2}{9}  (r^2 -\frac{1}{r}) \sin^2 \theta \ln(1+ \cos \theta).
\end{array}
\end{equation}
By using the boundary layer approximation
\begin{equation}
x= \frac{r-1}{\epsilon},
\end{equation}
the temperature field near interface can be expressed as
\begin{equation}
t(x,\theta)= 1+\frac{\pi}{6\sqrt{3}} -\frac{1}{6} \ln48
+\frac{2}{3} \ln (\frac{1 +\cos \theta}{\sin \theta}) + \frac{1}{3} x \sin^2 \theta \epsilon
\ln \epsilon + o(\epsilon \ln \epsilon ).
\end{equation}

\subsection{Outer temperature field within the drop}

By using the outer expansion for the scaled temperature
field within the droplet in Eqs. (4)
\begin{equation}
T' = \frac{1}{\epsilon^2}T'_{-2} +\frac{1}{\epsilon}T'_{-1} + T'_0 + o(1),
\end{equation}
the equation in its leading order can be written as
\begin{equation}
\begin{array}{l}
u' \frac{\partial{T'_{-2}}}{\partial{r}} +\frac{v'}{r} \frac{\partial{T'_{-2}}}{\partial{\theta}}= 0.
\end{array}
\end{equation}
Its solution is
\begin{equation}
\begin{array}{l}
T'_{-2} = F_0(\psi'),
\end{array}
\end{equation}
where $\psi'=\frac{3}{4} \sin^2 \theta (r^4 - r^2)$ is the
streamfunction within the droplet. The unknown function $F_0(\psi')$
can be obtained from the following equation for the temperature
field $T'_0$ in its second order
\begin{equation}
1+ u' \frac{\partial{T'_0}}{\partial{r}} +\frac{v'}{r} \frac{\partial{T'_0}}{\partial{\theta}}=\lambda \Delta F_0.
\end{equation}

Following \cite{7}, in the solving Eq. (25), the coordinate transformation from $(r,
\theta)$ to $(m,q)$ is applied in the form of
\begin{equation}
\begin{array}{l}
m = -\frac{16}{3} \psi',\\
q = \frac{r^4 \cos^4 \theta}{2r^2-1},
\end{array}
\end{equation}
where $m$ and $q$ denote the streamlines and their orthogonal lines, respectively.
The solution of Eq.(25) is thus written as
\begin{equation}
T'_{-2}(r, \theta)= F_0= \frac{1}{\lambda} [B' - \frac{1}{16}m + \frac{3}{256}(3\ln2-1\frac{3}{4})m^2 -\frac{3}{512}m^2\ln m] + o(m^2\ln m),
\end{equation}
where $B'$ is an unknown constant.
By using the boundary layer approximation
\begin{equation}
x'= \frac{1-r}{\sqrt{\lambda}\epsilon},
\end{equation}
the temperature field near interface can be expressed as follows
\begin{equation}
\begin{array}{ll}
t'(x', \theta) & =
-\frac{1}{2\sqrt{\lambda}}  x' \sin^2 \theta  \frac{1}{\epsilon}
- \frac{3}{8} x'^2 \sin^4  \theta  \ln \epsilon
+ o(\ln \epsilon).
\end{array}
\end{equation}

\subsection{ Inner temperature fields in the leading order}

By using inner expansions for the continuous phase and the fluid
in the droplet
\begin{eqnarray}
t(x, \theta)= t_{-1} \frac{1}{\epsilon}  + t_{l0} \ln \epsilon
+ o(\ln \epsilon ),\\
t'(x', \theta)=  t'_{-1} \frac{1}{\epsilon} + t'_{l0} \ln \epsilon
 + o(\ln \epsilon )
\end{eqnarray}
and the inner variables given in Eqs. (20) and (28), the scaled
energy equations for the  inner temperature fields in the leading
order can be written as follows
\begin{eqnarray}
-3x \cos \theta \frac{\partial t_{-1}}{\partial x} + \frac{3}{2} \sin \theta \frac{\partial t_{-1}}{\partial \theta}
= \frac{\partial^2 t_{-1}}{\partial x^2},\\
-3x' \cos \theta \frac{\partial t'_{-1}}{\partial x'} + \frac{3}{2} \sin \theta \frac{\partial t'_{-1}}{\partial \theta}
= \frac{\partial^2 t'_{-1}}{\partial x'^2}.
\end{eqnarray}
The boundary conditions are
\begin{equation}
\begin{array}{l}
t_{-1}(0,\theta) =t'_{-1}(0,\theta),\\
\delta \frac{\partial{t_{-1}}}{\partial x}(0,\theta) + \omega \delta \sin^2 \theta \cos \theta = -\frac{\partial{t'_{-1}}}{\partial x'}(0,\theta),\ \ \ 0 \leq \theta \leq \pi/2,\\
\delta \frac{\partial{t_{-1}}}{\partial x}(0,\theta) = -\frac{\partial{t'_{-1}}}{\partial x'}(0,\theta),\ \ \ \pi/2<\theta \leq \pi,\\
t_{-1}(x \to \infty, \theta) \to 0,\\
t'_{-1}(x' \to \infty, \theta) \to B - \frac{1}{2 \sqrt{\lambda}} x' \sin^2 \theta,
\end{array}
\end{equation}
where $\delta=\sqrt{\lambda}/\beta$ and $\omega =\Omega \epsilon^2=\frac{2}{3}(1+\frac{2\beta}{\lambda})$. We transform the independent variables from $[(x,x'),\theta]$ to
$[(\eta,\eta'),\xi]$ and the functions from $(t_{-1},t'_{-1})$ to
$(f_0,f'_0)$ as
\begin{equation}
\begin{array}{l}
(\eta,\eta')= (\frac{3}{2} x \sin^2 \theta, \frac{3}{2} x' \sin^2 \theta),\\
\xi=\frac{1}{2}(2-3\cos \theta +\cos^3 \theta) = \frac{1}{2} (2+ \cos \theta)(1- \cos \theta)^2
\end{array}
\end{equation}
and
\begin{equation}
\begin{array}{l}
f_0(\eta,\xi)= t_{-1}(x, \theta),\\
f'_0(\eta',\xi)= t'_{-1}(x', \theta) - B  + \frac{1}{2 \sqrt{\lambda}} x' \sin^2 \theta.
\end{array}
\end{equation}
The corresponding energy equations for $f_0,f'_0$ and the boundary
conditions can be written as follows
\begin{equation}
\begin{array}{l}
\frac{\partial f_0}{\partial \xi} =\frac{\partial^2 f_0}{\partial \eta^2},\\
\frac{\partial f'_0}{\partial \xi} =\frac{\partial^2 f'_0}{\partial \eta'^2}\\
\end{array}
\end{equation}
and
\begin{equation}
\begin{array}{l}
f_0(0, \xi)= f'_0(0,\xi) +B,\\
\delta \frac{\partial{f_0}}{\partial \eta}(0,\xi)
= -\frac{\partial{f'_0}}{\partial \eta'}(0,\xi) + \frac{1}{3\sqrt{\lambda}} + \Phi(\xi),\ \ \ 0 \leq \xi \leq 1,\\
\delta \frac{\partial{f_0}}{\partial \eta}(0,\xi) = -\frac{\partial{f'_0}}{\partial \eta'}(0,\xi) +\frac{1}{3\sqrt{\lambda}},\ \ \ \ 1 < \xi \leq 2,\\
f_0(\eta \to \infty, \xi)= 0,\\
f'_0(\eta' \to \infty, \xi) =0,
\end{array}
\end{equation}
where $\Phi(\xi) = - \frac{2 \omega \delta}{3} \cos \theta = - \frac{2 \omega \delta}{3}
(\cos \frac{\phi}{3} -\sqrt{3} \sin \frac{\phi}{3}), \phi=\arccos (1-\xi)$ in the Shengjin's formula \cite{18}.
To solve Eqs. (37), initial conditions are provided below
\begin{equation}
\begin{array}{l}
f_0(\eta,0) =0,\\
f'_0(\eta',0) =f'_0(\eta',\xi(\pi)) =f'_0(\eta',2) =g_0(\eta'),\\
g_0(\eta' \rightarrow \infty) \rightarrow 0.
\end{array}
\end{equation}
Following the methods given by Carslaw \& Jaeger \cite{16} and Harper \& Moore \cite{17}, the solutions
of Eqs.(37) for the continuous phase
and the fluid in the droplet can be respectively determined as
\begin{equation}
\begin{array}{ll}
f_0(\eta,\xi) =& \frac{1}{1+\delta} \{- \frac{1}{\sqrt{\pi}} \int_0^\xi \Phi(\xi-\tau) \exp(-\frac{\eta^2}{4\tau}) \frac{d \tau}{\tau^{1/2}} + (B +\frac{\eta}{3\sqrt{\lambda}}) {\rm erfc}(\frac{\eta}{2\sqrt{\xi}}) \\
& + \frac{1}{\sqrt{\pi\xi}} \int_0^{\infty} g_0(\eta^*) \exp[-\frac{(\eta +\eta^*)^2}{4\xi}]d \eta^* \}, (0 \leq \xi \leq 1), \\
f_0(\eta,\xi) =& \frac{1}{1+\delta} \{- \frac{1}{\sqrt{\pi}} \int_0^\xi \Phi(\xi-\tau) \exp(-\frac{\eta^2}{4\tau}) \frac{d \tau}{\tau^{1/2}}
 + \frac{1}{\sqrt{\pi}} \int_0^{\xi-1} \Phi(\xi-1-\tau) \exp(-\frac{\eta^2}{4\tau}) \frac{d \tau}{\tau^{1/2}}\\
& + (B+ \frac{\eta}{3\sqrt{\lambda}}) {\rm erfc}(\frac{\eta}{2\sqrt{\xi}})
 + \frac{1}{\sqrt{\pi\xi}} \int_0^{\infty} g_0(\eta^*) \exp[-\frac{(\eta +\eta^*)^2}{4\xi}]d \eta^* \}, (1 < \xi \leq 2) \\
\end{array}
\end{equation}
and
\begin{equation}
\begin{array}{ll}
f'_0(\eta',\xi) =& \frac{\delta}{1+\delta} \{ - \frac{1}{\sqrt{\pi}\delta} \int_0^\xi \Phi(\xi-\tau) \exp(-\frac{\eta'^2}{4\tau}) \frac{d \tau}{\tau^{1/2}}
 -(B - \frac{\eta'}{3\delta\sqrt{\lambda}}) {\rm erfc}(\frac{\eta'}{2\sqrt{\xi}}) \} \\
&+ \frac{1}{2\sqrt{\pi \xi}} \int_0^{\infty} g_0(\eta^*) \{ \exp[-\frac{(\eta'-\eta^*)^2}{4\xi}]
+ \frac{1-\delta}{1+\delta} \exp[-\frac{(\eta'+\eta^*)^2}{4\xi} ] \}d\eta^* , (0 \leq \xi \leq 1),\\
f'_0(\eta',\xi) =& \frac{\delta}{1+\delta} \{ - \frac{1}{\sqrt{\pi}\delta} \int_0^\xi \Phi(\xi-\tau) \exp(-\frac{\eta'^2}{4\tau}) \frac{d \tau}{\tau^{1/2}}\\
& + \frac{1}{\sqrt{\pi}\delta} \int_0^{\xi-1} \Phi(\xi-1-\tau) \exp(-\frac{\eta'^2}{4\tau}) \frac{d \tau}{\tau^{1/2}}
 -(B - \frac{\eta'}{3\delta\sqrt{\lambda}}) {\rm erfc}(\frac{\eta'}{2\sqrt{\xi}}) \} \\
&+ \frac{1}{2\sqrt{\pi \xi}} \int_0^{\infty} g_0(\eta^*) \{ \exp[-\frac{(\eta'-\eta^*)^2}{4\xi}]
+ \frac{1-\delta}{1+\delta} \exp[-\frac{(\eta'+\eta^*)^2}{4\xi} ] \}d\eta^* , (1<\xi \leq 2).\\
\end{array}
\end{equation}

\subsection{Steady migration velocity of the droplet}

Due to the zero net force acting on the droplet at the flow direction,
the migration speed of the droplet can be obtained as
\begin{equation}
V_\infty = -\frac{1}{2(2+3\alpha)} \int_0^{\pi} \sin^2 \theta \frac{\partial t}{\partial \theta} (0,\theta) d\theta
 = \frac{1}{2+3\alpha} \int_0^{\pi} \sin \theta \cos \theta t (0,\theta) d\theta.
\end{equation}
When the inner expansion in the temperature field (30) is truncated
at the $o(\ln \epsilon)$ order,  we rewrite Eq. (42) as
\begin{equation}
V_\infty = \frac{1}{2+3\alpha} \int_0^{\pi} \sin \theta \cos \theta [t_{-1} (0,\theta) \frac{1}{\epsilon}  +t_{l0} (0,\theta) \ln \epsilon] d\theta.
\end{equation}
Since $\epsilon=1/\sqrt{MaV_{\infty}}$, the migration speed of the droplet is evaluated as
\begin{equation}
V_\infty \approx a_1^2 Ma -2 a_{l0} \ln Ma + a_0,
\end{equation}
where
\begin{equation}
a_1 = \frac{1}{2+3\alpha} \int_0^{\pi} \sin \theta \cos \theta t_{-1} (0,\theta) d\theta
\end{equation}
and
\begin{equation}
a_{l0} = \frac{1}{2+3\alpha} \int_0^{\pi} \sin \theta \cos \theta t_{l0} (0,\theta) d\theta.
\end{equation}
From Eqs.(40), we obtain the  inner
temperature field in its leading order for the continuous phase near the surface of the droplet
\begin{equation}
\begin{array}{rl}
t_{-1}(0, \theta)=& f_0(0,\xi) \\
=& \frac{1}{1+\delta} [- \frac{1}{\sqrt{\pi}} \int_0^\xi \Phi(\xi-\tau) \frac{d \tau}{\tau^{1/2}}+ B+  \frac{1}{\sqrt{\pi \xi}} \int_0^{\infty} g_0(\eta^*) \exp(-\frac{{\eta^*}^2}{4\xi})d\eta^* ]\\
=& \frac{1}{1+\delta} [- \frac{1}{\sqrt{\pi}} \int_0^{\sqrt{\xi}} \Phi(\xi-s^2) d s + B+  \frac{2}{\sqrt{\pi}} \int_0^{\infty} g_0(2 \xi^{1/2} \zeta) \exp(-\zeta^2)d\zeta], 0 \leq \theta \leq \pi/2,\\
t_{-1}(0, \theta)
=& f_0(0,\xi) \\
=& \frac{1}{1+\delta} [- \frac{1}{\sqrt{\pi}} \int_0^\xi \Phi(\xi-\tau) \frac{d \tau}{\tau^{1/2}}+ \frac{1}{\sqrt{\pi}} \int_0^{\xi-1} \Phi(\xi-1-\tau) \frac{d \tau}{\tau^{1/2}} +B\\
&+ \frac{1}{\sqrt{\pi \xi}} \int_0^{\infty} g_0(\eta^*) \exp(-\frac{{\eta^*}^2}{4\xi})d\eta^* ]\\
=& \frac{1}{1+\delta} [- \frac{1}{\sqrt{\pi}} \int_0^{\sqrt{\xi}} \Phi(\xi-s^2) d s + \frac{1}{\sqrt{\pi}} \int_0^{\sqrt{\xi-1}} \Phi(\xi-1-s^2) d s + B\\
&+  \frac{2}{\sqrt{\pi}} \int_0^{\infty} g_0(2 \xi^{1/2} \zeta) \exp(-\zeta^2)d\zeta], \pi/2 < \theta \leq \pi.
\end{array}
\end{equation}
Substituting Eq. (47) into Eq. (45), we obtain
\begin{equation}
\begin{array}{ll}
a_1 =&-\frac{1}{\sqrt{\pi}(2+3\alpha)(1+\delta)} \{ \int_0^\pi \sin \theta
\cos \theta [\int_0^{\sqrt{\xi}} \Phi(\xi-s^2) d s] d \theta\\
& - \int_{\pi/2}^{\pi} \sin \theta
\cos \theta [\int_0^{\sqrt{\xi-1}} \Phi(\xi-1-s^2) d s]d \theta \}\\
& +\frac{2}{\sqrt{\pi}(2+3\alpha)(1+\delta)} \int_0^{\pi} \sin \theta \cos \theta [\int_0^{\infty}
g_0(2 \xi^{1/2} \zeta) \exp(-\zeta^2)d\zeta] d\theta.
\end{array}
\end{equation}
To determine the function $g_0$ in Eq. (48), we use the boundary
condition within the droplet at the front and rear stagnation points
in Eq. (39)
\begin{equation}
\begin{array}{ll}
g_0(\eta') =& \frac{\delta}{1+\delta} \{ - \frac{1}{\sqrt{\pi}\delta} \int_0^{\sqrt{2}} \Phi(2-s^2) \exp(-\frac{\eta'^2}{4s^2}) d s\\
& + \frac{1}{\sqrt{\pi}\delta} \int_0^1 \Phi(1-s^2) \exp(-\frac{\eta'^2}{4s^2}) d s
 -(B -\frac{\eta'}{2\delta \sqrt{\lambda}}) {\rm erfc}(\frac{\eta'}{2\sqrt{2}}) \} \\
&+ \frac{1}{2\sqrt{2\pi}} \int_0^{\infty} g_0(\eta^*) \{ \exp[-\frac{(\eta'-\eta^*)^2}{8}]
+ \frac{1-\delta}{1+\delta} \exp[-\frac{(\eta'+\eta^*)^2}{8} ] \}d\eta^*.\\
\end{array}
\end{equation}
The integral of the fourth
term on the right-hand side of Eq. (49) is approximated as
\begin{equation}
\int_0^\infty g_0(\eta^*) h(\eta',\eta^*) d\eta^* = \int_0^{\eta^*_l}g_0(\eta^*) h(\eta',\eta^*) d\eta^*
+ g_0(\eta^*_l) \int_{\eta^*_l}^{\infty} h(\eta',\eta^*) d\eta^*.
\end{equation}
 Then, Eq. (49) is evaluated in a linear system of equations
\begin{equation}
\begin{array}{ll}
& g_0(\eta') - \frac{1}{4\sqrt{2\pi}} g_0(\eta^*_1)  \{\exp[-\frac{(\eta'-\eta^*_1)^2}{8}]
+ \frac{1-\delta}{1+\delta} \exp[-\frac{(\eta'+\eta^*_1)^2}{8}] \} \Delta \eta^*\\
& - \frac{1}{4\sqrt{2\pi}} g_0(\eta^*_{N+1})  \{\exp[-\frac{(\eta'-\eta^*_{N+1})^2}{8}]
+ \frac{1-\delta}{1+\delta} \exp[-\frac{(\eta'+\eta^*_{N+1})^2}{8}] \} \Delta \eta^*\\
& - \frac{1}{2\sqrt{2\pi}} \sum_{j=2}^N g_0(\eta^*_j)  \{\exp[-\frac{(\eta'-\eta^*_j)^2}{8}]
+ \frac{1-\delta}{1+\delta} \exp[-\frac{(\eta'+\eta^*_j)^2}{8}] \} \Delta \eta^*\\
& - \frac{1}{2} g_0(\eta^*_{N+1}) [ {\rm erfc} (\frac{\eta^*_{N+1}-\eta'}{2\sqrt{2}})
+  \frac{1-\delta}{1+\delta} {\rm erfc} (\frac{\eta^*_{N+1}+\eta'}{2\sqrt{2}}) ]\\
=& \frac{\delta}{1+\delta} [- \frac{1}{\sqrt{\pi}\delta} \int_0^{\sqrt{2}} \Phi(2-s^2) \exp(-\frac{\eta'^2}{4s^2}) d s\\
& + \frac{1}{\sqrt{\pi}\delta} \int_0^1 \Phi(1-s^2) \exp(-\frac{\eta'^2}{4s^2}) d s
-(B -\frac{\eta'}{2\delta \sqrt{\lambda}}){\rm erfc} (\frac{\eta'}{2\sqrt{2}}) ],\\
\end{array}
\end{equation}
where $\eta^*_{N+1}=\eta^*_l$ and $\Delta \eta^*= \eta^*_l/N$.
The physical coefficients used in space experiments \cite{13} with the uniform temperature gradient
$G=12$ K/cm for the continuous phase of Fluorinert
FC-75 and the droplet of 5cst silicon oil at $T=333$ K are adopted to yield
$\alpha=0.342$, $\beta=0.571$ and $\lambda=0.299$.
A typical value for $\eta^*_l$ is chosen as 3.
Using the trial
and error method to satisfy the above approximation, we determine
the unknown constant $B=1.419$ and
obtain the dependence of $g_0$ on $\eta'$ as shown in Fig. 3.
From Eq. (48), we can determine the root-mean-square of the leading order term of the migration
speed as
\begin{equation}
a_1 =4.354 \times 10^{-2}.
\end{equation}
Although equations and boundary conditions describing the second order term of the migration speed
can be obtained, we are unable to find an analytical result for $t_{l0}$ in Eq. (46).
Under the truncation after the leading order term in Eq.(44), we obtain the migration speed of the droplet
\begin{equation}
V_\infty \approx 1.896 \times 10^{-3} Ma,
\end{equation}
which indicates that the thermocapillary droplet migration speed
increases as Ma number increases.
Using the migration speed $V_\infty$, Eq. (15) is rewritten as
\begin{equation}
\begin{array}{l}
\Omega = \frac{2}{3} (1 + \frac{2 \beta}{\lambda}) V_\infty Ma \approx 6.098 \times 10^{-3} Ma^2.
\end{array}
\end{equation}
Therefore, to reach steady thermocapillary droplet migration in the space experiment at large Ma numbers \cite{13},
the external radiation energy source
\begin{equation}
\begin{array}{l}
S=\Omega \sin^2 \theta \approx 6.098 \times 10^{-3} Ma^2 \sin^2 \theta
\end{array}
\end{equation}
  in contrary to the direction of movement should be provided.

\section{Conclusions and discussions}
\label{sec:sum} In this paper, the steady thermocapillary droplet migration in a
uniform temperature gradient combined with a radiation energy source at large Re and Ma numbers is studied.
The magnitude of the radiation energy source with the sine square dependence is determined to preserve
the conservative integral thermal flux across the surface. Under the assumption
 of quasi-steady state, we have determined an analytical result for the steady thermocapillary
migration of droplet at large Re and Ma
numbers. The result shows that the thermocapillary droplet migration
speed increases with the increasing of Ma number.

In general, when the droplet in a uniform temperature gradient moves upward, the thermal energy
is not only transferred into the droplet from the top surface but also out the droplet from the bottom surface.
Meanwhile, the thermal flux across the surface is balanced.
For large Ma numbers, once thermocapillary droplet migration reaches a quasi-steady state, the relation
of the nonconservative integral thermal flux across the surface
will be required \cite{11,12}. To satisfy the challenge, a thermal source at the surface through the absorption from
an external radiation energy source is provided for the system to make a balance of the integral thermal flux
across the surface.
The thermal source at the surface can bring more heat
 to the droplet, while the heat transfer in the system due to the thermal conduction across/around the droplet is weaker than that due to the thermal convection around the droplet at large Ma numbers.
 The thermocapillary migration of a droplet in the uniform temperature gradient combined with the radiation energy source at large Ma numbers can thus arrive at a quasi-steady state process.

To perform a real space experiment to confirm the above theoretical analysis
of the steady thermocapillary migration of a droplet,
the laser beam heating technology may be one of the possible
physical means to provide the external radiation energy source in contrary to the droplet movement direction.

\newpage
\textbf{Acknowledgments} This research is supported by the National Natural Science Foundation
of China through the Grants No. 11172310 and No. 11472284.
The author thanks the IMECH research computing facility for
assisting in the computation.

\newpage
\textbf{Figure caption}

 Fig.~1. A schematic diagram of the thermocapillary droplet migration under the combined actions of
 a temperature gradient $G$ and a radiation energy source $S=\Omega \sin^2 \theta$ in an axisymmetric spherical coordinate system moving with the droplet velocity $V_\infty$.

 Fig.~2. A schematic diagram of potential flows and boundary layer flows of
 the thermocapillary droplet migration at large Re numbers. Solid line: the interface of the droplet;
 Dashed/Longdash lines: the interface between potential flow (white/green zone) and boundary layer flow (blue/yellow zone) in the continuous flow/within the droplet;
 DashDot lines: streamlines of the potential flows inside and outside the droplet.

 Fig.~3. Function $g_0$ versus $\eta'$ determined from Eq. (51).

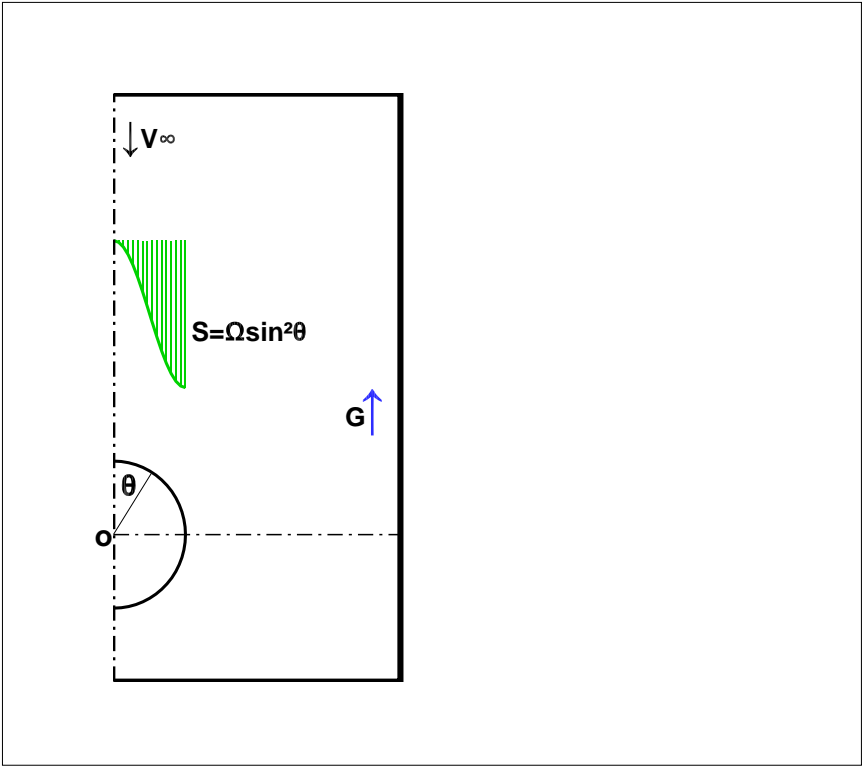
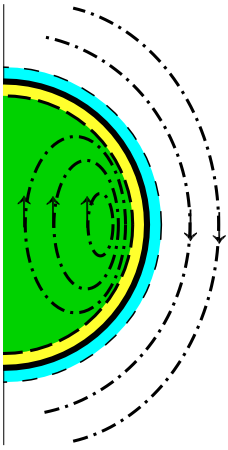
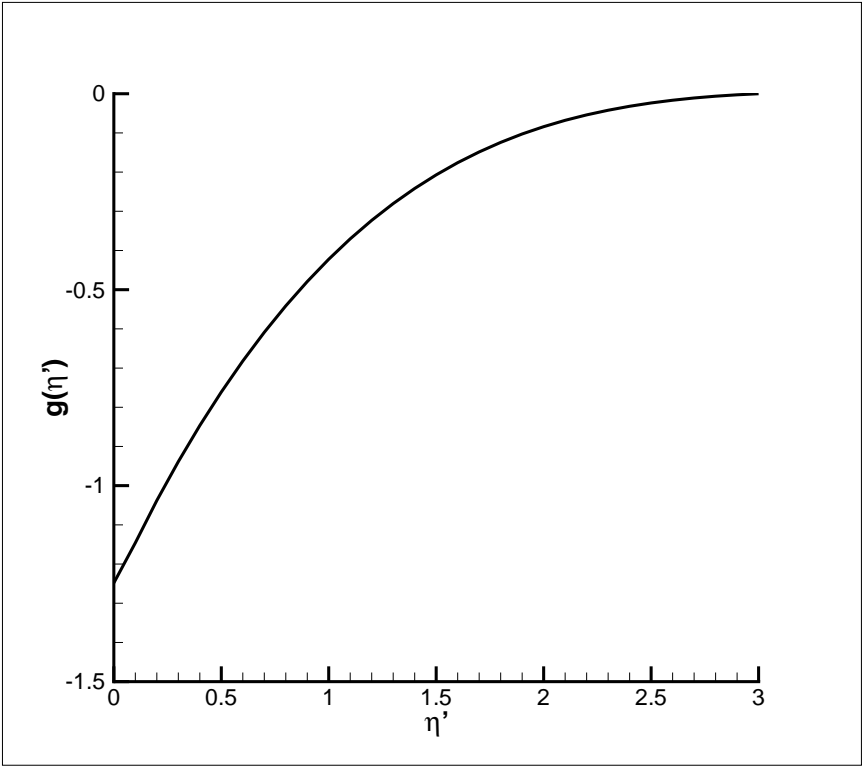
\end{document}